\documentclass[a4paper,11pt]{article}
\pdfoutput=1
\usepackage{jinstpub}
\usepackage[T1]{fontenc}
\usepackage{graphicx}
\usepackage{booktabs}
\usepackage{siunitx}

\title{An end-to-end experimental pipeline for automated track identification for paleo-detectors}

\author[a,1]{Claudio~Galelli,\note{Corresponding author.}}
\author[a,b]{Alessandro~Testi,}
\author[a]{and Lorenzo~Caccianiga}

\affiliation[a]{INFN sezione di Milano,\\via Celoria 16, 20133 Milano, Italy}
\affiliation[b]{Dipartimento di Fisica, Universit\`a degli Studi di Milano, \\via Celoria 16, 20133 Milano, Italy}

\emailAdd{claudio.galelli@mi.infn.it}

\abstract{
Paleo-detectors aim to exploit the damage tracks left in natural minerals by cosmic messengers over geological times to reconstruct past fluxes of energetic particles. For cosmic rays, the expected signal is abundant, and the tracks are micron-scale, so they can be revealed by etching and read out with conventional optical microscopy; what remains challenging is turning a bulk sample into a calibrated track-length spectrum that can be quantitatively confronted with theoretical predictions. In this work we present the experimental pipeline we developed for this purpose within the INFN \texttt{PRImuS} experiment, from sample preparation and automated multi-focus microscopy to track recognition with \texttt{OptimusPrimus}, a two-stage machine-learning algorithm, and to a forward-folding model that propagates geometric sectioning, etching and the measured detection response into simulated spectra. The pipeline was built and validated on chemically etched obsidian where the recovered track-length distribution is statistically compatible with expert annotation. In parallel, we developed and characterized a dry argon-plasma etching protocol for halite, an evaporite mineral of particular interest as a low-background cosmic-ray paleo-detector. Since every stage of the pipeline is mineral-agnostic, with only the trained networks and the detection response to be re-derived from annotated data, we are now applying it to plasma-etched halite with the aim of identifying cosmic-ray-induced tracks.
}

\keywords{Detector design and construction technologies and materials; Solid state detectors; Detector modelling and simulations I; Analysis and statistical methods; Radiation calculations; Image filtering; Pattern recognition, cluster finding, calibration and fitting methods; Radiation damage evaluation methods}

\begin{document}
\maketitle

\flushbottom

\section{Introduction}
\label{sec:intro}

When an energetic particle traverses and interacts within a crystal lattice, it transfers kinetic energy to a primary knock-on atom. If this energy exceeds the threshold displacement energy of the lattice (typically tens of eV), the atom is permanently displaced from its structural position and initiates a cascade of subsequent collisions, generating a dense trail of structural defects. In insulating minerals, the disruption prevents immediate recombination of the defects, leaving a permanent, non-crystalline damage zone known as a latent track. The best-known application of this effect is fission-track and alpha-recoil dating, in which the tracks are produced by the energetic offspring of spontaneous fission and $\alpha$-decay of radioactive isotopes in the sample and are used to infer its radiogenic content or its age \cite{Fleischer1975,Wagner1992}. The novel and developing paleo-detector idea instead targets the tracks left in natural minerals by the external fluxes of energetic particles that constantly bombard the Earth's crust, coming from the atmosphere, the Solar System, and the Galaxy and Universe at large \cite{Baum2023}.

The experimental demands of this technique depend strongly on the signal being sought. Paleo-detectors were originally proposed for rare-event physics, such as WIMP dark-matter interactions and coherent elastic neutrino-nucleus scattering \cite{Drukier2019,Baum2023}, where the rarity of the interaction is the key feature, since an old and clean mineral offers an enormous passively accumulated exposure (mass $\times$ time) that no artificial real-time detector could match. The nuclear recoils of interest are, however, rare and low-energy, producing latent tracks in the \SIrange{1}{100}{nm} range, and the target must be exceptionally radiopure and shielded from the cosmic-ray flux, since any radiogenic or cosmogenic background would generate indistinguishable tracks in the same length range. A track of a few nm is far below the resolution of conventional optical microscopy, so this regime requires expensive, low-throughput nm-resolution readout such as helium-ion microscopy, small-angle X-ray scattering, atomic-force or transmission electron microscopy, or a radically different approach such as selective-plane illumination microscopy \cite{Vladimirov2024}.

Paleo-detectors for cosmic rays (CRs), the focus of this work, sit in a more forgiving regime. The steady dose of CR-induced nuclear recoils, which is a background for rare-event searches, becomes the signal itself \cite{Galelli2026gamma}. Because every mineral is continuously exposed to a CR intensity set by its overburden, the expected signal is abundant and dominated by energetic secondary recoils, yielding latent tracks in the micron-to-tens-of-micron range. A signal that is both abundant and optically resolvable does not require nanoscale instrumentation, but can be investigated with the etching-plus-optical-microscopy technique that has been used for decades to count fission tracks, which are of comparable size.

Exploiting this archive in practice still requires converting a bulk mineral sample into a calibrated track-length spectrum that can be compared with theoretical predictions. This is the goal of \texttt{PRImuS} (Paleo-astroparticles Reconstructed with the Interactions of MUons in Stone), an INFN-funded experimental proof of concept with two complementary technical objectives: (a) establishing argon-plasma etching as a more ecological, safer, and reliable alternative to wet chemical etching, in particular for minerals such as halite that cannot be etched in aqueous acid solutions; and (b) building a high-throughput optical microscopy and image-analysis workflow capable of scanning statistically meaningful sample areas \cite{Galelli2026gamma}. On the theory side, the expected track-length spectrum for a given mineral, flux history, and burial history is computed with a dedicated simulation chain \cite{primus-paleo} that couples a CR flux propagation module \cite{Fedynitch2015} to a Geant4 nuclear-recoil and ion range simulation \cite{Geant4}.

The two objectives have been developed at different speeds. Developing a reliable plasma etching protocol for halite is genuinely new ground: there is little literature on the kinetic etching of evaporites \cite{LaVoieIngram2023}, and the recipe has to be established empirically, together with an understanding of the plasma source itself. Building the image-analysis and comparison chain, on the other hand, does not need to wait for that protocol, and is best validated on a sample whose tracks are unambiguous and whose etching is well understood. We therefore pursued the two in parallel. The automated track recognition and the forward-folding comparison were developed and validated on chemically etched volcanic obsidian containing fission tracks, a well-known solid-state nuclear track detector (SSNTD) in which the expected morphology of tracks is understood, while the halite plasma-etching protocol was being characterized.

This paper documents the resulting pipeline end-to-end and its validation on obsidian, and lays out how it is now being transferred to halite, where the goal is the identification of CR-induced tracks. The central design choice that makes this transfer possible is that no component of the chain is specific to one type of sample: the microscopy acquisition and the network architectures are mineral-agnostic, and the only mineral-specific inputs, the trained network weights and the measured detection response, can be re-derived
from annotated images of the target mineral using the same procedures described here. Section~\ref{sec:prep} describes sample preparation for the obsidian benchmark and for halite. Section~\ref{sec:microscopy} describes the automated microscopy acquisition. Section~\ref{sec:optimus} describes \texttt{OptimusPrimus}, our two-stage track-recognition pipeline, and its performance on the obsidian benchmark. Section~\ref{sec:folding} describes how theoretical spectra are folded with geometric, etching, and detection effects so that they can be compared with a measured distribution, and section~\ref{sec:conclusion} discusses the transfer to halite and the remaining limitations.

\section{Sample preparation}
\label{sec:prep}

\subsection{Obsidian validation benchmark}
\label{sec:obsidian}

Obsidian, a natural volcanic glass, is the calibration target within \texttt{PRImuS}. Its fission tracks, produced by the spontaneous fission of ${}^{238}$U, have been studied for decades in the context of fission-track dating \cite{Fleischer1975,Wagner1992}: their etching behaviour, morphology, and length scale are well documented, and in an optically clean glass, essentially every etch pit of the expected shape is a genuine track. This makes obsidian an optimal testbed for the readout and recognition stages, as any failure of the pipeline can be attributed to the pipeline itself rather than to an immature sample-preparation protocol or an ambiguous signal, and the micron-scale fission tracks fall in the same size range as the CR-induced tracks expected in paleo-detectors.

\begin{figure}[htbp]
\centering
\includegraphics[width=0.6\textwidth]{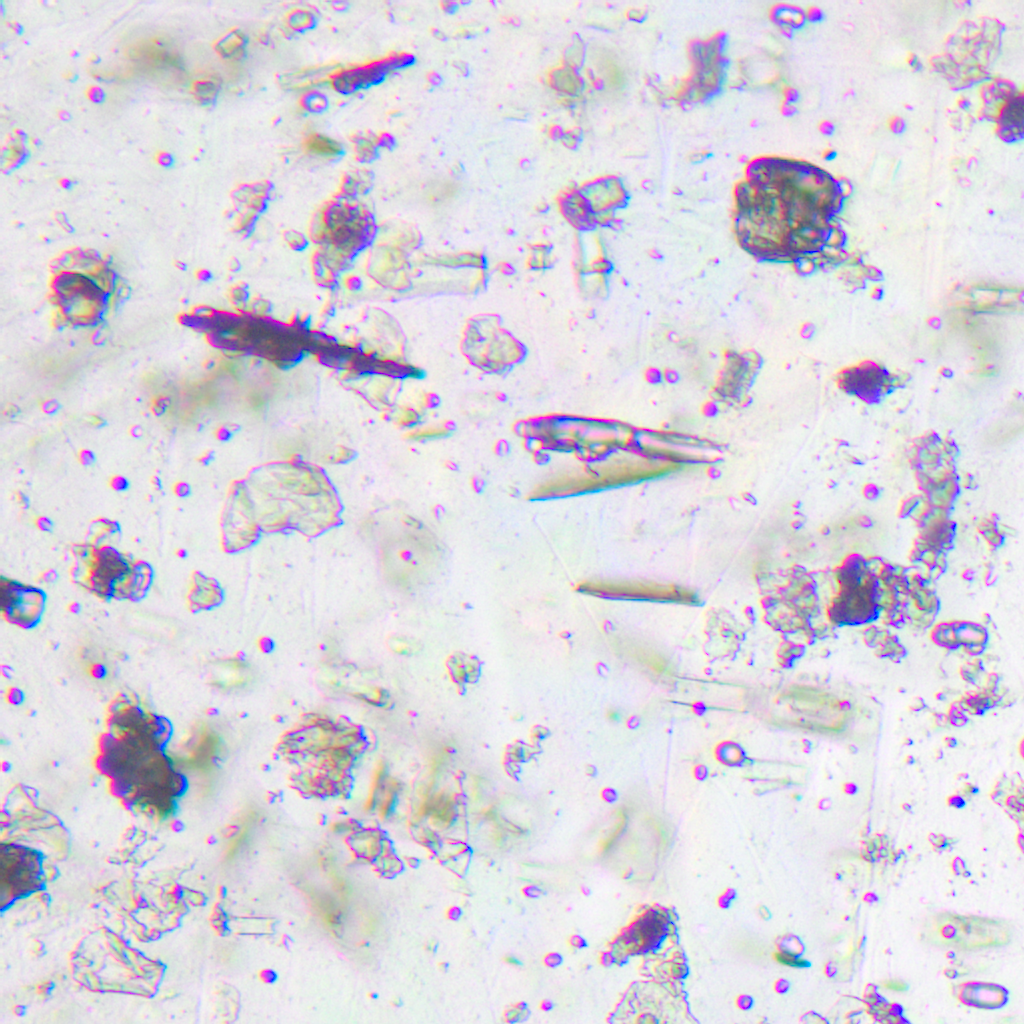}
\caption{Chemically etched volcanic obsidian. The elongated drop-like etch pits are fission tracks.}
\label{fig:obsidian}
\end{figure}

The obsidian samples, from the Monte Arci volcanic complex (Sardinia), were provided by the fission-track laboratory of the Physics Department of the Universit\`a degli Studi di Milano, where the preparation protocol and track identification criteria used here were learned. Some of the samples were irradiated with slow neutrons to induce fission in ${}^{235}$U, and increase the density of tracks; these samples were used for model training (section~\ref{sec:datasets}), while the remaining samples were used for later tests of the detection model (section~\ref{sec:folding}). Monte Arci is one of the four Mediterranean obsidian sources exploited in prehistory, and its flows have been extensively characterized for provenance studies \cite{Tykot1997,Meloni2002,Luglie2006,Bonizzoni2023}. Samples are encased in resin for handling stability, mechanically polished with diamond paste to also obtain a flatter surface, and etched in \SI{20}{\percent} hydrofluoric (HF) acid at \SI{40}{\celsius} for \SI{120}{s}. This is a standard route for revealing fission tracks in glassy targets \cite{FleischerPrice1964glass,Fleischer1975,Wagner1992}, routinely used for the fission-track dating of Mediterranean archaeological obsidians \cite{BigazziBonadonna1973,Bigazzi2005,Martinelli2020,Bonizzoni2022,Martinelli2025}, and was adopted without modification so that the benchmark would rest on an established procedure. A representative etched field is shown in figure~\ref{fig:obsidian}.

\subsection{Halite}
\label{sec:halite}

Halite (NaCl) is an evaporite of particular interest for \texttt{PRImuS} and the paleo-detector community in general because of its low intrinsic radioactivity (uranium content as low as \SI{e-10}{g/g}), which keeps the radiogenic track background under control over the long exposure windows relevant to CR flux reconstruction. In our studies, Messinian halite, deposited during the Mediterranean salinity crisis, has been identified as a possible probe of the CR flux some \SI{6}{Myr} ago, and the expected track-length spectra have already been computed for a range of flux scenarios \cite{Caccianiga2024}. Unlike obsidian, halite is highly water-soluble and therefore not amenable to aqueous acid etching; its behaviour under kinetic plasma etching has sparked interest in the mineral detection community \cite{LaVoieIngram2023}. Establishing a stable plasma etching recipe for halite is accordingly one of the two core technical goals of \texttt{PRImuS} \cite{Galelli2026gamma}.

Beyond being compatible with a water-soluble target, argon-plasma etching has advantages that motivate its adoption across \texttt{PRImuS} more broadly. Chemical etching of most SSNTD minerals relies on hazardous reagents, primarily HF, that demand dedicated handling protocols and protective equipment and produce a corrosive liquid waste stream. A dry plasma process removes this hazard and its waste entirely, making the technique safer and less polluting for high-throughput operation. Because argon is chemically inert, the etching mechanism is physical sputtering rather than a chemically selective dissolution: whereas an acid can preferentially attack specific crystallographic planes or defect chemistries, an inert plasma acts on the sample nearly isotropically and is expected to enlarge damage tracks with little directional bias, regardless of their orientation relative to the etched surface.

Halite specimens in \texttt{PRImuS} lab tests are cleaved from a bulk crystal along parallel cleavage planes, so that the analysed surface is a fresh, as flat as possible cleavage face, and are glued to a glass slide to ensure consistent handling and mounting on the microscope. Etching is performed in a Diener Atto low-pressure plasma system with an MHz-frequency generator (nominal power \SIrange{0}{300}{W}), operated at full power with argon as the process gas. After the chamber is evacuated, argon is introduced, and the pressure is allowed to stabilize at a readout pressure of \SI{0.2}{mbar} for several minutes before ignition; no deliberate change of pressure or flow is made during treatment. The reference protocol is a \SI{45}{min} etch at \SI{0.2}{mbar}. The laboratory where the experimental setup is located experiences a relative humidity of about \SI{40}{\percent}, and as such all samples are kept in dry storage containers with desiccant silica gel.

\begin{figure}[htbp]
\centering
\setlength{\tabcolsep}{2pt}
\renewcommand{\arraystretch}{0.8}
\begin{tabular}{@{}r cccc@{}}
 & {\small Pristine} & {\small \SI{15}{min}} & {\small \SI{45}{min}} & {\small \SI{135}{min}} \\[2pt]
\rotatebox{90}{\small\hspace{1.2em}$10\times$} &
\includegraphics[width=0.23\textwidth]{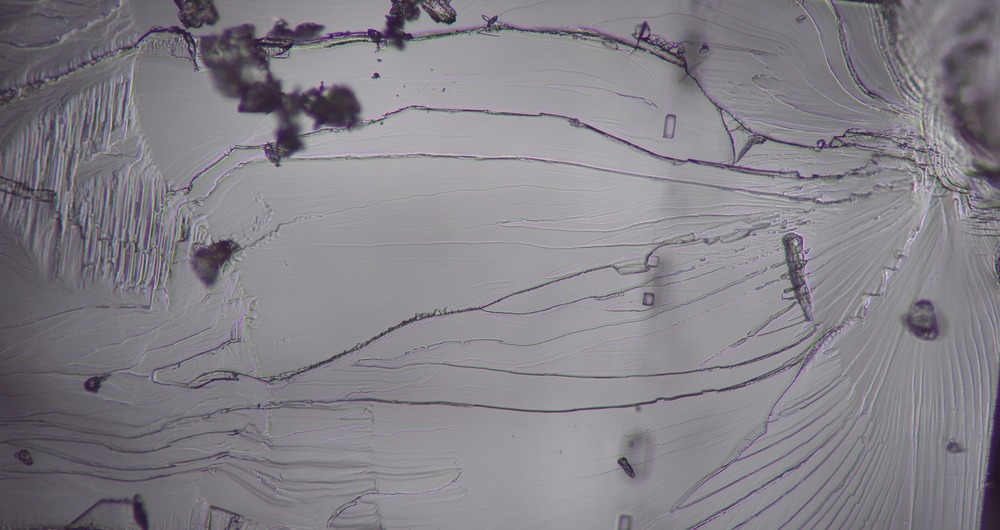} &
\includegraphics[width=0.23\textwidth]{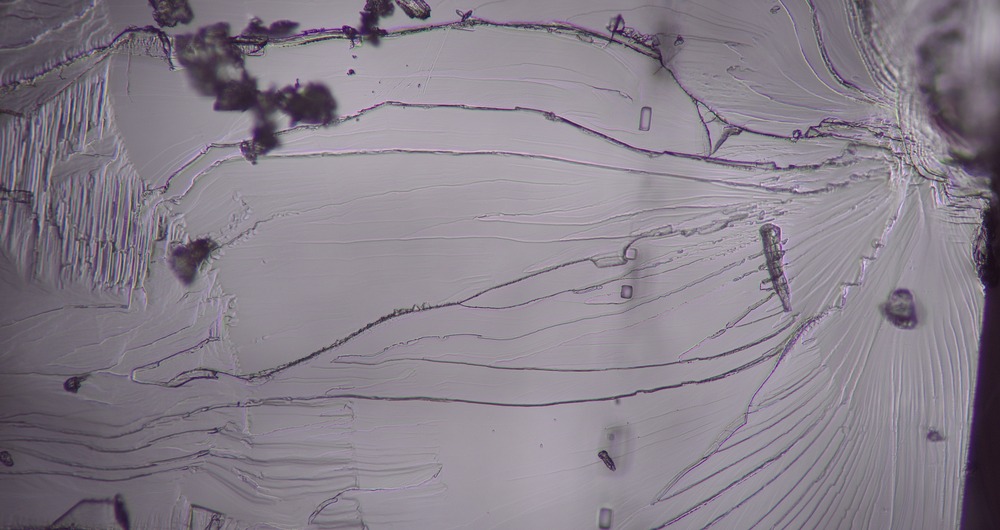} &
\includegraphics[width=0.23\textwidth]{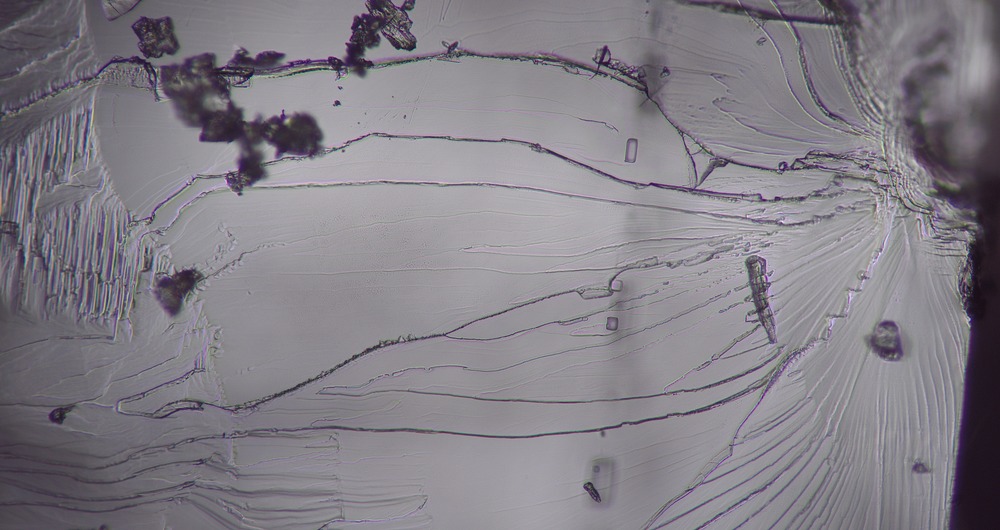} &
\includegraphics[width=0.23\textwidth]{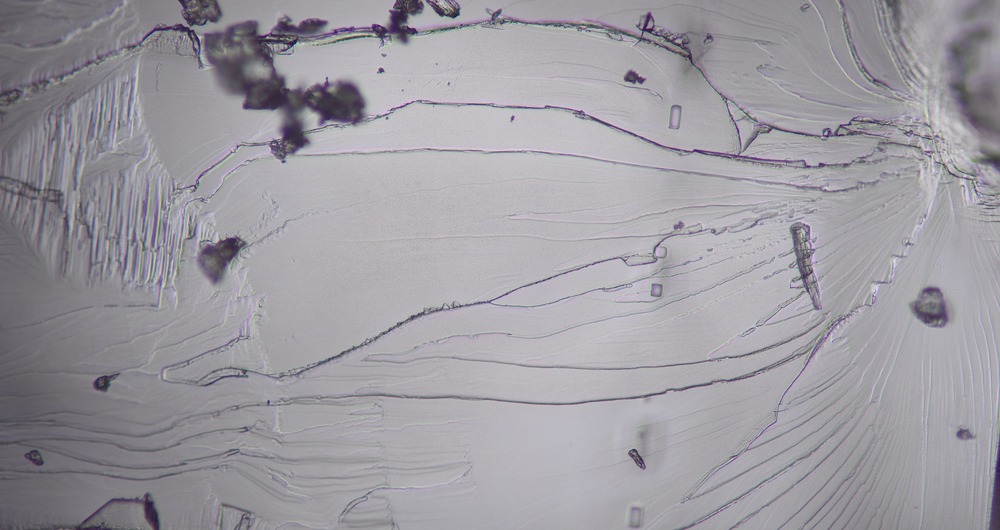} \\
\rotatebox{90}{\small\hspace{1.2em}$40\times$} &
\includegraphics[width=0.23\textwidth]{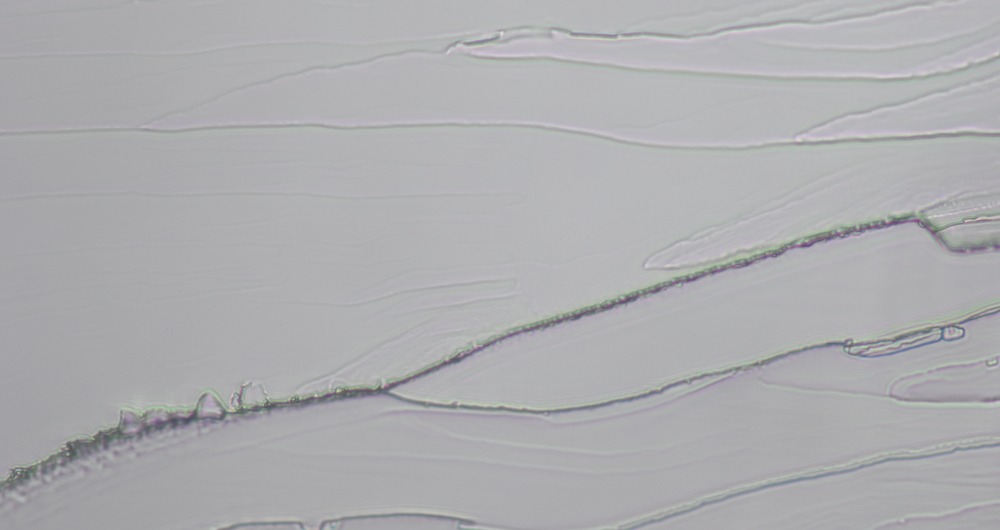} &
\includegraphics[width=0.23\textwidth]{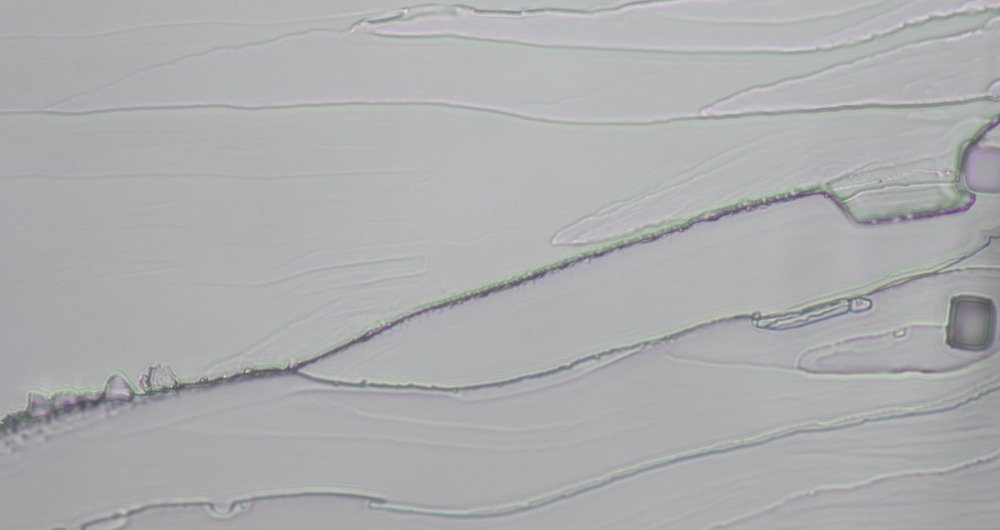} &
\includegraphics[width=0.23\textwidth]{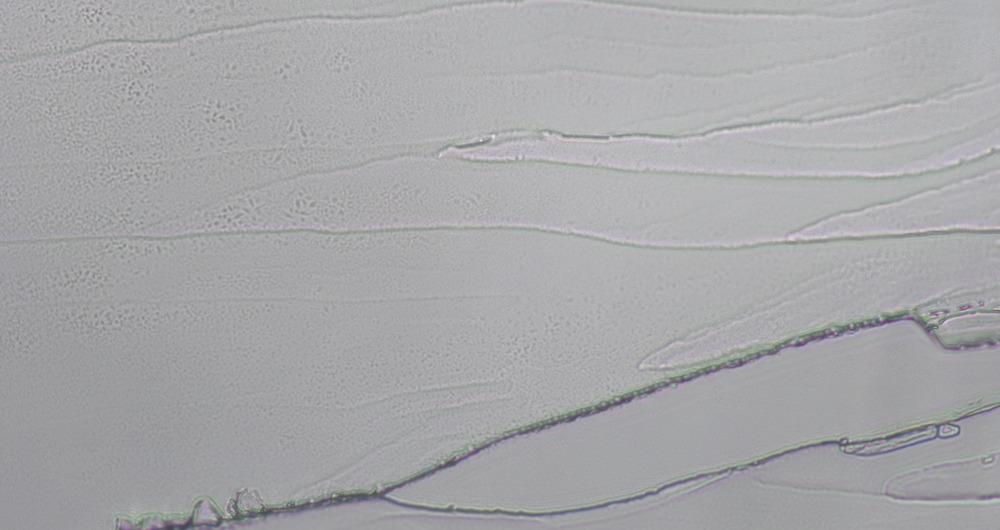} &
\includegraphics[width=0.23\textwidth]{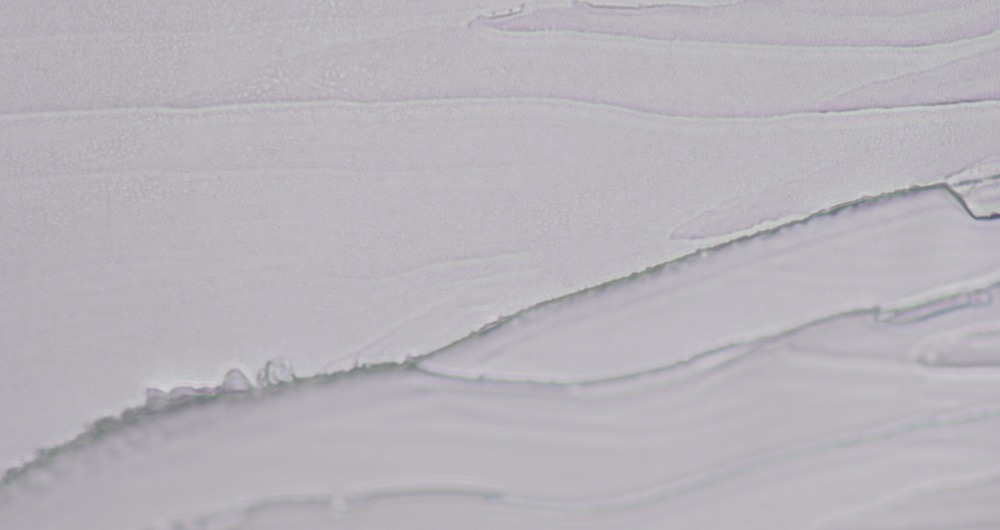} \\
\end{tabular}
\caption{The same cleaved halite surface imaged before treatment (left) and after cumulative argon-plasma exposures of 15, 45, and \SI{135}{min} at \SI{0.2}{mbar}, at $10\times$ (top row) and $40\times$ (bottom row) magnification. The progressive roughening of the surface demonstrates the action of the plasma. Scale bars: \SI{50}{\micro m} (top), \SI{10}{\micro m} (bottom).}
\label{fig:halite_etch}
\end{figure}

Having established that the plasma treatment acts on the halite surface, as shown by the progressive roughening of the cleave face with increasing exposure (figure~\ref{fig:halite_etch}), we are now acquiring large-area image sets of plasma-etched halite at the reference conditions. These will be used both to search for cosmic-ray-induced etch pits and to build the annotated dataset needed to retrain \texttt{OptimusPrimus} on this mineral (section~\ref{sec:conclusion}). The same field, imaged after increasing cumulative exposures, also motivates the choice of a \SI{45}{min} reference etch. After \SI{15}{min} the surface is only mildly modified. By \SI{45}{min} the roughening is clearly developing but not overwhelming. After \SI{135}{min} the surface texture becomes dense enough to compete with any small feature in the optical images, which would make track candidates harder to separate from the background. A \SI{45}{min} etch therefore sits at the time edge of where the plasma roughens the surface clearly while keeping the surface clean enough for optical readout.

\section{Microscopy readout}
\label{sec:microscopy}

Etched samples are imaged with an Evident (formerly Olympus) BX53MTRF-mot upright optical microscope with a motorized $x$-$y$-$z$ stage, operated in transmission with a $10\times$ apochromatic objective and an 8-megapixel dedicated CMOS sensor, giving a pixel scale of \SI{0.345}{\micro m}; $40\times$ and $100\times$ objectives are available for detailed inspection of individual features. Acquisition is handled by the Evident-Olympus proprietary PRECiV software. For each sample, the region of interest is defined by tracing its contour once on a rough scan map of the whole slide; the software then tiles this contour into a grid of fields of view. A focus map across the tiles is computed automatically and checked manually, especially for rougher samples. At every field of view, a full focal stack is acquired rather than a single in-focus frame: 10 slices above and 20 below the nominal focus given by the focus map, with a \SI{12}{\micro m} increment, for 31 slices per field. This redundant sampling of the focal axis allows the image analysis pipeline (section~\ref{sec:optimus}) to select the sharpest available frames for a target that is not perfectly planar within a single depth of field, a common situation for a cleave surface with residual topography or a slight tilt.

The acquisition chain is identical for any sample: only the region of interest and the focus map change from sample to sample. The current end-to-end throughput is of order \SI{5}{cm^2} per day at 31 $z$-slices \cite{Galelli2026gamma}, limited mainly by the manual checks and the write-to-disk time of the full stack rather than by stage motion.

\section{The \texttt{OptimusPrimus} automated track recognition pipeline}
\label{sec:optimus}

Converting the image stacks into a list of individually measured tracks, and then into a track-length distribution, is the task of \texttt{OptimusPrimus}, a two-stage deep-learning pipeline. Its design decouples localization from verification: a semantic-segmentation network is tuned for high recall to generate candidate track regions, and a separate classification network then prunes false positives, since a single network optimized for both objectives is forced into a suboptimal compromise. For the \texttt{PRImuS} science case, missing genuine tracks is more costly than admitting a moderate rate of false candidates that a dedicated second stage can remove. The pipeline is implemented in PyTorch \cite{Paszke2019} and supports multi-GPU data-parallel training and inference, so that it scales with the throughput of the automated microscope.

The architecture, training procedure, and evaluation described in this section are generic. What is specific to a mineral are the annotated images used for training and, consequently, the network weights and the detection response derived from them (section~\ref{sec:folding}). All results reported here were obtained on the obsidian benchmark.

\subsection{Training and evaluation datasets}
\label{sec:datasets}

The segmentation network is trained on optical microscopy images for which pixel-wise binary masks of the track regions are produced by manual inspection. For the training run described here, we used chemically etched volcanic obsidian (section~\ref{sec:obsidian}) irradiated with slow neutrons to induce fission and increase the number of tracks. In total, 59 annotated images containing 2623 individual fission tracks were used, split into a training subset of 41 images (2047 tracks) and a validation/evaluation subset of 18 images (576 tracks). The latter serves both to tune the segmentation-mask binarization threshold and the classifier decision threshold, and to provide the performance figures of section~\ref{sec:performance}. For the annotated images, the focal plane used for training is selected manually.

The classification dataset is generated by combining automatic candidate extraction with manual review rather than from scratch. The trained segmentation model is applied to the available images, and each detected region is extracted as a $64\times64$-pixel patch (section~\ref{sec:classification}) and compared with the ground-truth mask to assign a track or background label. To increase variability beyond the 59 annotated images, the segmentation model can be further applied to unannotated $z$-stack images, and the resulting candidates reviewed and labelled manually before being added to the dataset. Alternatively, the classifier can be trained exclusively on a manually curated set of track and background patches; the two routes are kept separate rather than mixed. The same preprocessing and geometric normalization (section~\ref{sec:classification}) are applied during dataset generation, training, and inference, so that the classifier never encounters at inference time an orientation or intensity distribution that differs systematically from the training one.

For each acquired $z$-stack, images are first grouped by their spatial acquisition coordinates. The focus quality of each slice is scored with the variance of the image Laplacian, a standard blur-detection metric \cite{Bansal2016}, and only the sharpest slices are retained. This balances sensitivity to tracks that may be in best focus at different depths against the cost of processing the full 31-slice stack.

\subsection{Segmentation step}
\label{sec:segmentation}

The retained images are processed by an attention-based encoder-decoder segmentation network, a MAnet decoder \cite{Fan2020} on an EfficientNet-B7 encoder pretrained on ImageNet \cite{Tan2019}, which produces a per-pixel probability map of candidate track locations. Training uses a hybrid loss combining Dice and binary cross-entropy (BCE) in equal parts,
\begin{equation}
  \mathcal{L}_\mathrm{seg} = \tfrac12\,\mathcal{L}_\mathrm{Dice}
  + \tfrac12\,\mathcal{L}_\mathrm{BCE},
  \label{eq:segloss}
\end{equation}
optimized with AdamW \cite{Loshchilov2019} at an initial learning rate of $3\times10^{-4}$, a batch size of 2, and a ReduceLROnPlateau scheduler tied to the validation Dice score. Training images are augmented with random geometric transformations (flips, affine and elastic distortions) and intensity transformations (brightness and contrast variation, Gaussian noise, and CLAHE). Training runs for up to 200 epochs, with the checkpoint of the highest validation Dice retained; early stopping with a patience of 30 epochs is applied after a minimum of 100 epochs. After every epoch, the model is evaluated on the validation set with precision, recall, Dice, intersection-over-union (IoU), and the $F_2$ score, which weights recall more heavily than precision, consistent with the high-sensitivity role of this stage.

The probability maps of the retained slices are binarized with a threshold chosen to maximize the validation IoU rather than the validation precision, again favouring candidate inclusion over early rejection. Connected-component analysis of the aggregated mask yields a list of discrete candidate instances.

\subsection{Classification step}
\label{sec:classification}

Each candidate component defines a centered $64\times64$-pixel patch extracted from the source image, with symmetric zero-padding where the candidate lies too close to the border for a full crop. To remove orientation as a nuisance variable, an ellipse is fitted to each candidate and the patch is rotated so that its major axis is horizontal.

Normalized patches are classified by an EfficientNet-B0 network \cite{Tan2019}. The network is trained with a binary cross-entropy with logits loss, optimized with AdamW at an initial learning rate of $3\times10^{-5}$ and a weight decay of $10^{-3}$, with a ReduceLROnPlateau scheduler on the validation loss. A lighter augmentation than for segmentation is used (small affine transformations and Gaussian noise), since patches are already geometrically normalized. The checkpoint with the highest validation F1 is retained, and the decision threshold is optimized on the validation set by maximizing the $F_1$ score, in contrast to the recall-favouring operating point of the segmentation stage, consistent with the classifier's role as a verification step.

\begin{figure}[htbp]
\centering
\includegraphics[width=0.9\textwidth]{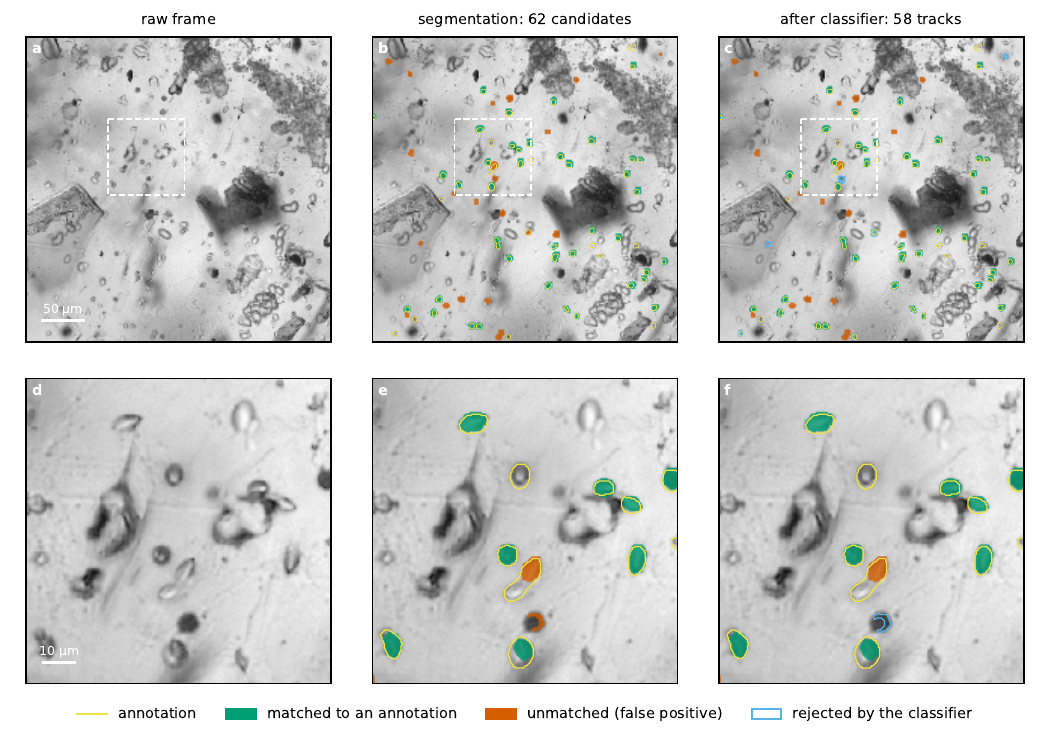}
\caption{Representative output of \texttt{OptimusPrimus} on an obsidian validation image (median F1 among the validation tiles). Top: full frame; bottom: the region marked by the dashed box. (a, d) Raw frame; (b, e) high-recall candidate mask from the segmentation stage; (c, f) tracks retained by the classifier. Detections are coloured by their match to the annotations (yellow outlines): matched (green), unmatched (orange); candidates rejected by the classifier are outlined in blue. Annotations with no detection inside are missed tracks.}
\label{fig:output}
\end{figure}

Finally, the validated components are fitted with ellipses to extract the centroid, the major- and minor-axis dimensions, the orientation, and a physical track length obtained from the calibrated pixel scale. These per-track measurements are exported as structured CSV files, from which per-sample track-length distributions are assembled. A representative output of the model on validation is shown in figure~\ref{fig:output}.

\subsection{Detection performance on the obsidian benchmark}
\label{sec:performance}

To quantify the effect of the classification stage, we compared the single-stage segmentation baseline with the full two-stage pipeline on the 18-image validation/evaluation set of section~\ref{sec:datasets}. Predicted and ground-truth instances are matched through an IoU criterion above a fixed threshold, solved as a global optimal assignment with the Hungarian algorithm \cite{Kuhn1955} rather than greedily in scan order, because a greedy match can let an early-processed prediction claim a suboptimal ground-truth instance and block a better-fitting prediction.

As shown in table~\ref{tab:detection}, the segmentation baseline reaches a recall $R=0.760$, locating the majority of true tracks ($N_\mathrm{TP}=438$), but optical clutter, surface scratches, and other surface artifacts produce a large number of false positives ($N_\mathrm{FP}=194$), for a precision $P=0.693$. Adding the classification stage reduces the false positives from 194 to 81 (a \SI{58.2}{\percent} reduction) at a cost in detected tracks ($N_\mathrm{TP}=354$, $R=0.615$), raising the precision to $P=0.814$. Because track-length investigation is more sensitive to false-positive contamination than to a moderate loss of recall, we report the $F_{0.5}$ score, which weights precision twice as heavily as recall: under this metric the full pipeline improves over the baseline ($0.706\to0.765$).

\begin{table}[htbp]
\centering
\caption{Track detection performance on the 18-image obsidian validation/evaluation set (576 annotated fission tracks).}
\label{tab:detection}
\begin{tabular}{lcccccc}
\toprule
Method & TP & FP & FN & Precision & Recall & $F_{0.5}$ \\
\midrule
Segmentation baseline & 438 & 194 & 138 & 0.693 & 0.760 & 0.706 \\
Full pipeline         & 354 &  81 & 222 & 0.814 & 0.615 & 0.765 \\
\bottomrule
\end{tabular}
\end{table}

Detection metrics alone do not establish whether the classifier removes tracks indiscriminately or preferentially removes artifacts, which is the property that matters for a length-distribution measurement. We therefore compared the predicted track lengths with the manual annotations using a two-sample Kolmogorov-Smirnov (K-S) test (table~\ref{tab:lengths}). The segmentation baseline yields $N=632$ objects with a systematic underestimation of the mean length ($\bar L = 5.32\pm1.34$\,\si{\micro m}, mean absolute error $\mathrm{MAE}=\SI{0.36}{\micro m}$), and the K-S test shows that this distribution departs significantly from the ground truth ($D=0.1107$, $p=0.0011$): the unfiltered artifacts admitted by the high-recall stage distort the recovered length profile. With the classification stage ($N=435$), the mean length error drops to \SI{0.16}{\micro m} ($\bar L=5.52\pm1.16$\,\si{\micro m}) and the K-S test gives $D=0.0675$, $p=0.1973$, so that the hypothesis of equal distributions is not rejected. The classifier thus preferentially removes artifact-driven false positives such as scratches, surface steps, and inclusions, rather than discarding candidates indiscriminately. The consequential loss of recall is real, but it is close to length-independent and is explicitly accounted for by the folding model of section~\ref{sec:folding}.

\begin{table}[htbp]
\centering
\caption{Track-length distributions on the obsidian validation set compared with the manual ground truth.}
\label{tab:lengths}
\begin{tabular}{lcccc}
\toprule
Method & $N$ & Mean $\pm$ std (\si{\micro m}) & MAE (\si{\micro m}) & K-S $p$ \\
\midrule
Ground truth      & 576 & $5.68\pm1.53$ & -   & -     \\
Segmentation only & 632 & $5.32\pm1.34$ & 0.36 & 0.0011 \\
Full pipeline     & 435 & $5.52\pm1.16$ & 0.16 & 0.1973 \\
\bottomrule
\end{tabular}
\end{table}

These results constitute the validation of the method: on a sample with a well-understood track population, \texttt{OptimusPrimus} recovers a track-length distribution that is statistically indistinguishable from expert annotation, with a measured detection response.

\section{Comparison with theoretical track length spectra}
\label{sec:folding}

The theoretical track-length spectrum produced by the simulation framework for a given mineral, flux history and burial history \cite{primus-paleo} is a volumetric quantity, the expected number of latent tracks per unit target mass, whereas a microscope image of an etched surface measures the openings of etch pits per unit area. A rigorous comparison must therefore account for the sectioning of the tracks by the analysed surface, for the way etching turns a latent track into a pit of a given size, and for the imperfect detection response of the recognition pipeline. We describe these steps in turn. The procedure is written for a generic mineral and etching technique.

\subsection{Geometric slicing and etching}
\label{sec:slicing}

Given the latent spectrum $n_i$, expressed as tracks per milligram of target in length bin $i$, we build a Monte Carlo realization of the track population sectioned by the analysed surface. The reference volume is a cube of mass $m=\SI{1}{mg}$, of side $a=(m/\rho)^{1/3}$ for a target of density $\rho$. A slab of thickness $a$ and surface $A$, equal to the analysed area, contains $A/a^2$ such cubes, and therefore
\begin{equation}
  N = \frac{A}{a^2}\sum_i n_i
  \label{eq:ntracks}
\end{equation}
tracks. Each synthetic track is assigned a latent length $L$, drawn uniformly within a bin chosen with probability proportional to $n_i$, so that wide theoretical bins do not produce a comb of discrete lengths. It also receives a dip angle $\phi$ with respect to the surface plane ($\phi=\pi/2$ for a track perpendicular to it), isotropic in solid angle by default, i.e.\ $\sin\phi$ uniform in $[0,1]$; a user-supplied angular density can be used instead. Finally, the depth $z_t$ of the upper end of the track below the original surface is drawn uniformly in $[-a/2,a/2]$; negative values describe tracks crossing the original surface, of which only the part below it exists.

Etching removes a bulk layer of thickness $h=v_B t$, where $v_B$ is the bulk etch rate and $t$ the etching time, and dissolves the damaged material along the track at the faster rate $v_T$, with $V=v_T/v_B>1$. A track can produce a pit only if the etchant reaches it, $z_t<h$, and if part of it lies below the original surface, $z_t+L\sin\phi>0$. The opening of the pit on the final surface is obtained by assuming constant etch rates. The etchant reaches the track at depth $z_s=\max(z_t,0)$, after which a bulk thickness $h'=h-z_s$ remains to be removed. It then travels along the track at $v_T$, over at most the latent length below that point, $R=(z_t+L\sin\phi-z_s)/\sin\phi$. A point at distance $s\le s_\mathrm{max}=\min(R,\,Vh')$ along the track is reached when a thickness $s/V$ of the remaining bulk layer has been removed, and from then on it dissolves outwards at the bulk rate. By the end of the etch it has grown by $r(s)=h'-s/V$ vertically and by $\lambda\,r(s)$ laterally, where $\lambda$ is the ratio of lateral to vertical bulk etch rate. Its distance from the final surface is $\Delta z(s)=h'-s\sin\phi$, so it contributes to opening a segment of half-width
\begin{equation}
  a(s)=\lambda\sqrt{r(s)^2-\Delta z(s)^2},
  \qquad r(s)>|\Delta z(s)|,
  \label{eq:huygens}
\end{equation}
centred at $x(s)=s\cos\phi$ along the projection of the track. The pit opening is the union of these segments, with major and minor axes
\begin{equation}
  D=\max\,[x(s)+a(s)]-\min\,[x(s)-a(s)],
  \qquad
  d=2\max a(s).
  \label{eq:axes}
\end{equation}
This construction covers the main regimes of track etching. Tracks with $\sin\phi<1/V$ never open, since $r(s)<\Delta z(s)$ at every point; this is the critical angle of the technique. Below it, it reproduces the conical phase of pit growth, the rounding of the pit tip once the etch front has reached the end of the track (over-etching), and the fading of shallow pits that the bulk etch has consumed. The sectioning bias of the technique follows automatically: longer and steeper tracks are more likely to be intersected and to produce large openings.

Pits whose minor axis is smaller than a resolution width $w_\mathrm{min}$, of about three pixels in our images, cannot be resolved optically and are discarded. For the remaining pits, the major axis $D$, which is the length reported by \texttt{OptimusPrimus}, is histogrammed directly on the bins used for the detection descriptor (section~\ref{sec:efficiency}). This gives the expected absolute number $N_i$ of pits measured on the area $A$ in each bin of etched size, without further normalization. The etching parameters $(h, V, \lambda)$ characterize the technique. For HF-etched obsidian we use $h=\SI{4}{\micro m}$, $V=2.5$ and $\lambda=1$. For plasma-etched halite the values remain to be determined from the exposure series of section~\ref{sec:halite}; the low bulk removal rate observed so far suggests $V_\mathrm{plasma}\gg V_\mathrm{HF}$, to be confirmed by measuring individual pits.

\subsection{Detection-response folding}
\label{sec:efficiency}

The sliced spectrum $N_i$ is converted into the histogram that \texttt{OptimusPrimus} is expected to report by a forward-folding model with three ingredients. All are measured on an annotated validation set of the target mineral, with the same instance matching used for table~\ref{tab:detection}. Together they form a binned detection descriptor that is produced at training time and stored with the trained models. For obsidian the descriptor is measured on the annotated set of induced fission tracks (section~\ref{sec:datasets}) and applied to natural fission tracks in an unirradiated sample, the two populations differing only in their density and not in the etched-pit morphology.

For each bin of true length, the recall $R_i=\mathrm{TP}_i/G_i$ is the fraction of the $G_i$ annotated tracks matched by a detection. This describes the length dependence of the combined segmentation and classification stages without assuming a functional form.

The matched pairs of annotated length $L$ and reported length $\hat L$ are modelled as
\begin{equation}
  \hat L = L + b(L) + \varepsilon,\quad
  \varepsilon\sim\mathcal{N}\!\left(0,\sigma^2(L)\right),\quad
  b(L)=b_0+b_1L,\quad
  \sigma^2(L)=\sigma_0^2+\sigma_1^2L^2,
  \label{eq:response}
\end{equation}
with coefficients obtained by least squares. The constant term $\sigma_0$ describes length-independent contributions such as pixel quantization and the blur of the pit edge, while $\sigma_1$ describes errors that grow with length. The resulting migration matrix,
\begin{equation}
  M_{ji} = \frac{1}{\Delta_i}\int_{e_i}^{e_{i+1}}\!\mathrm{d}L\,
  \left[\Phi\!\left(\frac{e_{j+1}-L-b(L)}{\sigma(L)}\right)
  -\Phi\!\left(\frac{e_j-L-b(L)}{\sigma(L)}\right)\right],
  \label{eq:migration}
\end{equation}
with $e_j$ the bin edges, $\Delta_i=e_{i+1}-e_i$, and $\Phi$ the standard normal cumulative distribution function, is the probability that a detected track of true bin $i$ is reported in bin $j$. Its columns sum to less than one because part of the migration falls outside the histogram range. Since the annotations measure the visible etched pit, $L$ corresponds directly to the major axis $D$ of eq.~\eqref{eq:axes}.

The precision $P_j=\mathrm{TP}_j/(\mathrm{TP}_j +\mathrm{FP}_j)$, binned in reported length, quantifies the contamination by artifacts such as scratches, surface steps, and inclusions. The number of such artifacts is set by the surface quality and not by the number of tracks, so a precision measured on one sample cannot be transferred to a sample with a different track density. We therefore convert the false positives into a rate per unit area, $\varphi_j=\mathrm{FP}_j/A_\mathrm{val}$, with $A_\mathrm{val}$ the total area of the validation images. This point is central to the transfer between minerals, as the track and background density of a halite, or any other mineral, sample is expected to be different from that of any other benchmark, and only an area-normalized background rate is a meaningful property of the readout.

For a measured area $A$, the expected number of reported tracks in bin $j$ is
\begin{equation}
  \mu_j = \sum_i M_{ji}\,R_i\,N_i + \varphi_j A.
  \label{eq:mu}
\end{equation}
For a sample with the same track density as the validation set, this is equivalent to $\mu_j=\sum_i M_{ji}R_iN_i/P_j$, i.e.\ to a recall-to-precision scaling, with the difference that recall is evaluated at the true length, precision at the reported length, and the false positives are not smeared as if they were tracks.

The reported counts $n_j$ are independent Poisson variables of mean $\mu_j$, since thinning and migration of a Poisson population and the addition of an independent Poisson background preserve the Poisson character.

\begin{figure}[htbp]
\centering
\includegraphics[width=0.9\textwidth]{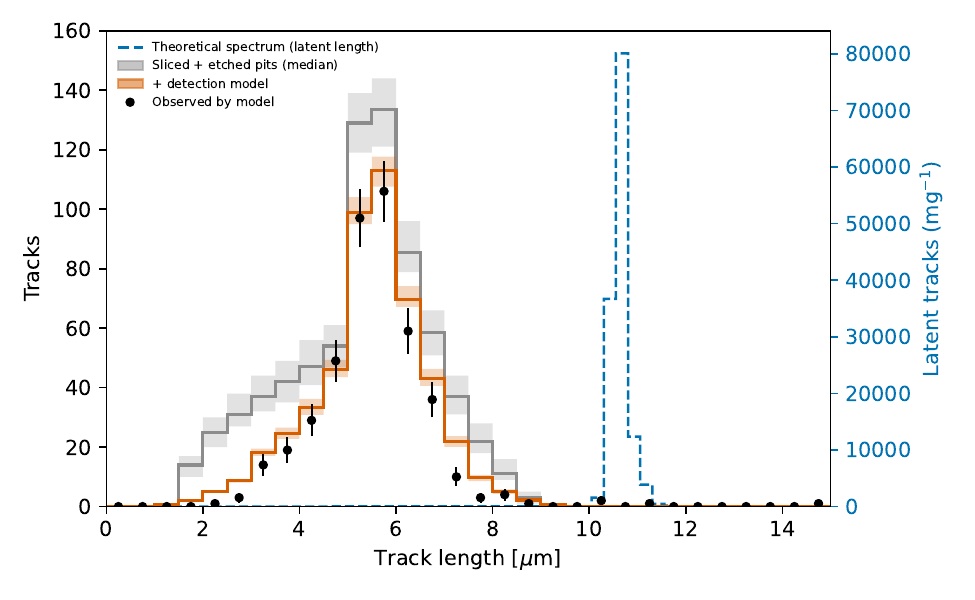}
\caption{Forward-folding chain applied to fission tracks in Monte Arci obsidian. Dashed blue line, right axis: theoretical latent track-length spectrum from ${}^{238}$U spontaneous fission, in tracks per mg. Grey: expected distribution of etched-pit major axes on the analysed area after sectioning and HF etching (eqs.~\eqref{eq:ntracks}--\eqref{eq:axes}). Orange: the same after folding with the detection response measured on the induced-track validation set (eq.~\eqref{eq:mu}). Black points: tracks reported by \texttt{OptimusPrimus}, with Poisson error bars. Shaded bands show the 1$\sigma$ spread over Monte Carlo realizations.}
\label{fig:obsidian_spectrum}
\end{figure}

Figure~\ref{fig:obsidian_spectrum} shows the full chain applied to natural fission tracks in Monte Arci obsidian. Sectioning and etching turn the narrow latent distribution, peaked around \SI{11}{\micro m}, into a broad distribution of pit openings peaked near \SI{6}{\micro m}, with a long tail of partially sectioned tracks towards small sizes. The detection response suppresses mainly this tail, where small pits are hardest to separate from surface features.

\section{Discussion and conclusion}
\label{sec:conclusion}

We have presented the experimental pipeline developed within \texttt{PRImuS} to turn a mineral sample into a track-length spectrum that can be compared quantitatively with the predictions of a cosmic-ray paleo-detector simulation. The pipeline was built and validated on Monte Arci obsidian, etched following an established fission-track dating protocol. On this benchmark, the two-stage \texttt{OptimusPrimus} recognition works as expected, with the classification step reducing false positives by \SI{58}{\percent} relative to segmentation alone, and returns a track-length distribution statistically compatible with expert annotation. The forward-folding model then propagates sectioning, etching, and the measured detection response into the theoretical prediction.

In parallel, we have characterized a dry etching protocol for halite. The reference recipe, \SI{45}{min} at \SI{0.2}{mbar}, avoids hazardous acids and liquid waste and is compatible with a water-soluble target. The progressive roughening of the cleave face with exposure time demonstrates that the treatment acts on the surface, and identifies \SI{45}{min} as the shortest tested exposure giving a clearly developed etch without the dense texture that appears in longer runs.

With the method validated on obsidian and the halite protocol established, we are now acquiring large-area image sets of plasma-etched halite, with the aim of identifying cosmic-ray-induced tracks. Because every stage of the pipeline was designed to be mineral-agnostic, this transfer does not require a new method but a new calibration. Halite images acquired with the reference protocol will be annotated following the same procedure adopted for obsidian. The segmentation and classification networks will then be retrained on these annotations, and the detection descriptor $\{R_i, M_{ji}, \varphi_j\}$ re-measured on an annotated halite validation set. The resulting spectra can finally be confronted, through eq.~\eqref{eq:mu}, with sample-specific track spectrum predictions computed for a range of cosmic-ray flux scenarios appropriate to the age of the sample, in an analogous way to the ones we published for Messinian Mediterranean halite \cite{Caccianiga2024}. In the longer term, the same chain is intended to be expanded to many interesting paleo-detector candidate minerals, such as volcanic olivine \cite{Galelli2026jcap}.

Several aspects will require particular attention in this transfer. The obsidian benchmark validates the method, not its halite-specific performance. Plasma-etched pits in a cubic crystal are expected to differ in morphology from HF-etched fission tracks in a glass, so recall and length response must be re-measured rather than assumed. The artifacts competing with genuine tracks are also mineral-specific. In halite, cleavage steps, dislocation etch pits, fluid-inclusion cavities and features related to the solubility of the surface replace the polishing scratches that dominate on obsidian. This makes the area-normalized false-positive rate of eq.~\eqref{eq:mu} the appropriate quantity to measure, and higher magnification imaging at $40\times$ and $100\times$ a valuable aid to annotation. The etching-specific inputs of the slicing model, $(h, V, \lambda)$, also have to be determined for the plasma protocol, from more exposure series and from higher-magnification imaging of individual pits. Finally, the throughput of the acquisition chain, currently limited by manual checks and data writing, will set the area, and hence the statistical reach, of the first halite measurements. This is the step that turns cosmic-ray paleo-detectors from a proposal into a measurement, and halite is where we are now taking it.

\acknowledgments
This work is supported by the Istituto Nazionale di Fisica Nucleare (INFN) CSN5 Grant for Young Researcher to the \texttt{PRImuS} project. We thank L.~Bonizzoni and A.~Guglielmetti (Universit\`a degli Studi di Milano) for providing the obsidian samples and for training us in sample preparation and fission-track identification. We also thank E.~LaVoie-Ingram (University of Michigan) for the help, fruitful discussions, and collaboration on argon plasma etching and mineral detection in general.

\bibliographystyle{JHEP}
\bibliography{refs}

\end{document}